\documentstyle[12pt]{article}
\input psfig.sty
\def\nn{\noindent}
\def\ie{{\it i.e.}}
\def\eg{{\it e.g.}}

\def\etal{{\it et al.}}

\def\to{\rightarrow}

\def\mpl{\ifmmode \overline M_{Pl}\else $\overline M_{Pl}$\fi}
\def\lsim{\mathrel{\mathpalette\atversim<}}

\renewcommand{\bar}[1]{\overline{#1}}

\def\grtsim{\,\,\rlap{\raise 3pt\hbox{$>$}}{\lower 3pt\hbox{$\sim$}}\,\,}
\def\lsim{\,\,\rlap{\raise 3pt\hbox{$<$}}{\lower 3pt\hbox{$\sim$}}\,\,}
\newcommand{\mP}{{\bar M}_P}
\newcommand{\lpi}{\Lambda_\pi}
\newcommand{\dcy}{G^{(3)} \to G^{(1)} G^{(1)}}

\begin{document}
\begin{flushright}
SLAC--PUB--8810\\
April 2001
\end{flushright}
\bigskip\bigskip

\thispagestyle{empty}
\flushbottom

\centerline{\Large\bf Bulk Physics at a Graviton Factory\footnote
{\baselineskip=14pt
Work supported by the Department of Energy, contract DE--AC03--76SF00515.}}
\vspace{22pt}

\centerline{\bf Hooman Davoudiasl and Thomas G. Rizzo}
\vspace{8pt}
  \centerline{\it Stanford Linear Accelerator Center}
  \centerline{\it Stanford University, Stanford, California 94309}
  \centerline{e-mail: hooman@slac.stanford.edu, rizzo@slac.stanford.edu}
\vspace*{0.9cm}

\begin{abstract}

A general prediction of the  5-d Randall-Sundrum (RS) hierarchy model is the emergence of
spin-2 Kaluza Klein (KK) gravitons with weak scale masses and couplings.  The lowest order
effective theory of the RS model is given by 5-d Einstein gravity which uniquely fixes the
self-interactions of gravitons.  We demonstrate that large numbers of light KK resonances could
be produced at a future lepton-collider-based ``Graviton Factory''.  Measuring the
self-interactions of these KK gravitons will probe the accuracy of the 5-d Einstein gravity picture
and, in addition, yield indirect information on the as yet untested self-coupling of the 4-d
graviton.  The self-interactions of the gravitons can be studied by measuring the decays of the
heavier states to the lighter ones.  Using the AdS/CFT picture, these can be interpretted as the
decays of heavier resonances to lighter ones, in a strongly coupled 4-d CFT.  We show that
these decays have sufficient rates to be studied at future colliders and that they are also, in
principle, sensitive to higher derivative operators, such as the Gauss-Bonnet term.  In a
generalized RS model, with non-universal 5-d fermion masses, FCNC's will be induced.  We
show that precise measurements of the rare decays of KK graviton/gauge states into flavor
non-diagonal final states at a Graviton Factory can be used to map the 5-d fermion mass matrix.  
     
\end{abstract}

\newpage

\section{Introduction}

The Randall-Sundrum (RS) model offers an interesting geometric view of 
the hierarchy between the
apparent scale of gravity $\mP \sim 10^{18}$ GeV and the weak scale 
$M_w \sim 10^2$ GeV \cite{rs}.  The RS
model background geometry is AdS$_5$ (5-d spacetime with constant negative 
curvature), truncated by
two 4-d Minkowski walls separated by a fixed distance.  All the parameters 
of the model are assumed to
be naturally given by various powers of the 5-d fundamental scale 
$M_5 \sim \mP$.  In the original RS
construct, the Standard Model (SM) fields are confined to one of the 4-d 
boundaries which we will refer to
as the SM wall.  The induced metric on the SM wall then generates from $M_5$ 
a physical scale $\lpi \sim
M_w$, through a geometric exponential warp factor.  Due to this 
exponentiation, no large hierarchies need
to be introduced; for more details on the RS construct please see 
Ref. \cite{rs}.  

A generic prediction of the RS model is the emergence of spin-2 resonances $G^{(n)}$, $n = 1, 2, 3,
\ldots$, which are the Kaluza-Klein (KK) excitations of the 5-d graviton.  The masses and couplings
of $G^{(n)}$ are set by $\lpi$, and hence, these resonances could be relevant to physics at the weak
scale.  Many recent works have studied the phenomenological effect of $G^{(n)}$ on collider and
precision electroweak data \cite{rsphen}.  These studies have focused on the leading coupling of
$G^{(n)}$ to the SM field content.  The RS model is based on an effective 5-d Einstein Gravity
(EG$_5$) theory which describes the coupling of the graviton to the SM fields, as well as its
self-couplings.  The $G^{(n)}$ are the 4-d manifestations of the 5-d graviton.  
Hence, to test the model at leading order, one must go beyond the $G^{(n)}-$SM
coupling and also probe the leading order inter-couplings of $G^{(n)}$ which are fixed, given $\lpi$
from collider data.  This is similar to testing QCD at tree level, which requires agreement between
theory and experiment, not only for the quark-gluon coupling, but also for the gluon self-couplings. 
The inter-couplings of $G^{(n)}$ can also yield indirect evidence for the as yet unobserved massless
4-d graviton self-coupling, as well as a handle on the coefficients of possible 5-d higher derivative
terms, such as the Gauss-Bonnet term.  

The original RS model has also been generalized to allow for the SM fields to reside in the 5-d bulk.  An
important set of parameters that enters this generalized RS model is the 5-d mass matrix of the SM bulk
fermions.  It has been shown that the value $\nu$ of the 5-d fermion mass in units of $k$, where $k$ is
the curvature scale of AdS$_5$, has a significant role in the low energy phenomenology of the model
\cite{rsfermions}, since it controls the couplings of the zero mode (observed) fermion to $G^{(n)}$ and the
KK gauge fields $A^{(n)}$.  It has been pointed out that giving various flavors of fermions different 5-d
masses induces FCNC's at tree level \cite{delA}.  Thus, these masses must be fairly universal in order to
avoid unacceptably large FCNC's.  One way of probing the 5-d fermion mass matrix is to search for
presumably rare flavor non-diagonal decays of $G^{(n)}$ and $A^{(n)}$.  

To probe graviton self-coupling or the 5-d masses of the fermions with precision in the
aforementioned manner, a large number of KK excitations are needed.  In this paper, we show that if a
generic RS-type model is realized in Nature an appropriate weak scale Linear Collider (LC) can be
used to produce large numbers, $\sim 10^7$, of light KK gravitons.  In this sense, the LC can
operate as a ``Graviton Factory''.  Of course, if the SM fields reside in the 5-d bulk, we expect to
produce comparably large numbers of light KK gauge fields as well.

As an example of the $G^{(n)}$ inter-coupling, we study the decay $G^{(3)} \to G^{(1)} G^{(1)}$, which is
the first kinematically allowed decay within the graviton sector.  We give the branching fraction for this
decay as a function of the ratio of the masses in the initial and final states; it is found that the dependence
on the mass ratio is fairly strong.  However, at the classical level in the RS model, the mass ratio is fixed
by the geometry and the branching fraction for $\dcy$ is about $15\%$.  These results are presented in
section 2, where brief comments on the 
AdS/CFT \cite{JM} interpretation of such decays and the 
effect of 5-d higher derivative terms are also included.

In section 3, we study a model in which all SM fermions except the top quark will be assumed to have the
same value of $\nu$.  We then show the effects of this mass matrix in terms of flavor non-diagonal
decays of $G^{(1)}$ and $A^{(1)}$.  Although this model is perhaps too simplistic to be
phenomenologically relevant, it demonstrates the utility of studying such decays at future Graviton
Factories in probing the bulk fermion  mass matrix.  Some concluding remarks are presented in section 4.  

\section{Graviton Sector Self-Interactions}

As mentioned before, the RS model generically predicts the emergence of spin-2 KK gravitons 
$G^{(n)}$ in the 4-d effective theory.  The equations of motion for $G^{(n)}$ are obtained through the 
KK reduction of the EG$_5$ action $S_G$, containing the 5-d graviton kinetic term.  We have 
\begin{equation}
S_G = 2 \, M_5^3 \int d^5x{ \sqrt {-G}} \, R_5,
\label{SG}
\end{equation}
where $G = det(G_{\mu \nu})$, $G_{\mu \nu}$ is the RS background metric, and $R_5$ is the 5-d Ricci 
scalar.  The 4-d interactions of the $G^{(n)}$ states with the SM fields are given by 
the following Lagrangian 
\begin{equation}
{\cal L}_{SM} = - \left[\frac{1}{\mP} h^{(0)}_{\mu \nu} + \frac{1}{\Lambda_\pi}
\sum_{n = 1}^{\infty} h^{(n)}_{\mu \nu}\right] T^{\mu \nu}, 
\label{LSM}
\end{equation}  
where the 4-d tensor fields $h^{(n)}_{\mu \nu}$ represent the $G^{(n)}$ states, and $T^{\mu \nu}$ is the
energy momentum tensor of SM which is assumed to reside on the SM wall, in Eq. (\ref{LSM}).  
Phenomenological studies of the RS model have so far focused on the interactions coming from 
${\cal L}_{SM}$, such as the on-shell decays of $G^{(n)}$ into SM fields.  These interactions are only 
suppressed by one power of $\Lambda_\pi$ for $n = 1, 2, 3, \ldots$.  

However, there are also interactions amongst $G^{(n)}$, resulting from the KK reduction of the
self-coupling of the 5-d graviton in $S_G$.  In powers of $M_5^{-1}$, the leading 5-d self-coupling is
the triple graviton vertex.  This indicates that in 4-d the $\{G^{(l)}, G^{(m)}, G^{(n)}\}$ coupling is, in powers of
$\Lambda_\pi^{-1}$, the leading interaction in the KK graviton sector.  This can be seen by 
using the KK graviton wavefunctions \cite{DHRprl}, and inspecting the triple KK graviton coupling of 
$S_G$.  Thus, to test the RS model to leading order in $\Lambda_\pi^{-1}$, one must also study the 
$\{G^{(l)}, G^{(m)}, G^{(n)}\}$ couplings.  We note that quartic couplings of the $G^{(n)}$ are higher order in 
$\Lambda_\pi^{-1}$, as they are for the 5-d graviton in powers of $M_5^{-1}$.  In principle, there could 
be higher derivative terms, such as $R_5^2$, in the 5-d theory.  However, 
their contribution to the $\{G^{(l)}, G^{(m)}, G^{(n)}\}$ couplings are higher order in $\Lambda_\pi^{-1}$.  We will 
briefly discuss the $R_5^2$ terms at the end of this section.

In the following, we study the process $\dcy$ which is the first kinematically allowed two body decay in the
KK sector.  To do this, we need the triple graviton vertex.  This vertex has the same tensor structure
regardless of the number of dimensions and the background geometry.  Hence, we use the Feynman rules
of Ref. \cite{BF} for the 3-graviton vertex $V_{\{\mu_i, \nu_i\}}(K_i), i = 1, 2, 3$, where $\{\mu_i, \nu_i\}$ are
4-d Lorentz indices and $K_i$ is the 4-momentum of the particle labeled $i$, in Ref. \cite{BF}.  Since
$V_{\{\mu_i, \nu_i\}}(K_i)$ has the same tensor structure in 5-d, we use the same function for the 5-d
graviton triple coupling, and treat $\{\mu_i, \nu_i\}$ and $K_i$ as 5-dimensional.  To get the decay rate for
$\dcy$, we must square $V_{\{\mu_i, \nu_i\}}(K_i)$ and contract it with the product of three massive
graviton polarization sums, to project out the appropriate degrees of freedom.  This sum is given in Ref.
\cite{HLZ} for the case of flat extra dimensions.  However, the degrees of freedom of the 4-d KK gravitons
are the same for the RS and the flat cases, independent of the higher dimensional theory.

The amplitude for the on-shell decay is obtained when the vertex is contracted with the 4-d massive
spin-2 polarization vectors of the KK states.  In this way, we see that the only relevant indices are 4-d and
we can take $\{\mu_i, \nu_i\}$ to be 4-d Minkowskian.  However, the scalar products of the 5-momenta,
$K_i \cdot K_j$, in $V_{\{\mu_i, \nu_i\}}(K_i)$ must be treated carefully.  By using the 4-d massive graviton
polarization sum we have extracted the appropriate tensor structure, but to complete the reduction from
5-d to 4-d, these scalar products should be expressed in terms of 4-d quantities.  The 5-momenta
$K_i^M, M = 0, 1, \ldots, 4$, in the Feynman rules correspond to 5-d derivatives $\partial^M$, acting on
the graviton field $h_i$.  Thus, $K_i \cdot K_j$ corresponds to terms proportional to
$G^{M N} \partial_M h_i \partial_N h_j$, in the 5-d action.  

The 5-d graviton can be expanded in the KK modes $h^{(n)}_{\mu \nu}$ 
\begin{equation} 
h_{\mu \nu}(x, \phi) =
\sum_{n = 0}^{\infty} h^{(n)}_{\mu \nu}(x) \, \frac{e^{-2\sigma}\chi^{(n)}(\phi)}{\sqrt{r_c}}, 
\label{KKh} 
\end{equation} 
where the
$Z_2$ even wavefunctions $\chi^{(n)}(\phi)$ only depend on the 5$^{th}$ dimension parameterized by $\phi \in
[-\pi, \pi]$; $x^4 = r_c \phi$, $\sigma = k r_c |\phi|$, 
and $r_c \sim 10 \, k^{-1}$ is the compactification radius.  Note that, in Eq. (\ref{KKh}), 
the graviton is taken to be a 5-d tensor fluctuation on the AdS$_5$ background.  
The 4-d 
derivatives $\partial_\mu$ do not act on $e^{-2\sigma}\chi^{(n)}(\phi)$ 
and thus, after integration over the extra dimension, the 4-d
part of the dot product $K_i \cdot K_j$ is an overall constant times the 4-d Minkowski scalar 
product, for a given vertex with fixed KK fields.  
However,
$\partial_\phi$ acts on $e^{-2\sigma}\chi^{(n)}(\phi)$ and, hence, the 5-d part of $K_i \cdot K_j$ 
depends on the mode number
$n$ of the KK states whose momenta are $K_i$ and $K_j$.  Therefore, for particles 
$\{G^{(m)}_i, G^{(n)}_j\}$, we
may schematically write 
\begin{equation} 
\int d\phi \, K_i \cdot K_j \to \alpha
 \, k_i \cdot k_j - \beta_{i j}(m, n), 
\label{KiKj} 
\end{equation}
where $k_i$ are the Minkowski 4-momenta.  We again note that $\alpha$ does not depend on the mode 
numbers $(m, n)$ of the particles whose
momenta are in the product, but $\beta (m, n)$ does.  That is, for the 4-d parts of the vertex, 
such as the terms
proportional to $k_i \cdot k_j$, $\alpha$ is the same and is given by integrating the wavefunctions
$\chi^{(n)}(\phi)$ for the particles in the vertex, say $\{G^{(l)}, G^{(m)}, G^{(n)}\}$, over the extra dimension. 
However, $\beta (m, n)$ are given by integrating a product proportional to 
$\chi^{(l)} \, \partial_\phi \chi^{(m)} \,
\partial_\phi \chi^{(n)}$, which depends on the modes $(m, n)$.

Using the notation of Ref. \cite{DHRprl}, we have      
\begin{equation}
\chi^{(n)} \simeq \frac{e^{2 \sigma}}{N_n} J_2 (z_n), 
\label{chi}
\end{equation}
where $J_l$ is the $l^{th}$ order Bessel function.  
The normalization is given by
\begin{equation}
N_n \simeq \frac{e^{k r_c \pi}}{\sqrt{k r_c}} J_2(x_n),  
\label{Nn}
\end{equation}
where $J_1(x_n) = 0$; $z_n = (m_n/k) e^{\sigma}$, 
and $m_n$ is the mass of the $n^{th}$ KK 
graviton.  Given $\chi^{(n)}$ from Eq. (\ref{chi}), integration of the $\{G^{(l)}, G^{(m)}, 
G^{(n)}\}$ vertex in $S_G$ of Eq. (\ref{SG}) over the extra dimension yields
\begin{equation}
\alpha = \lambda_1 \int_{-\pi}^{\pi} d\phi \, \, e^{4 \sigma} J_2 (z_l) J_2 (z_m) J_2 (z_n), 
\label{alpha}
\end{equation}
where 
\begin{equation}
\lambda_1 = \frac{(2 \mP)^{-1} k r_c \, e^{-3 k r_c \pi}}{J_2 (x_l) J_2 (x_m) J_2 (x_n)}.
\label{lambda1}
\end{equation}
We also obtain
\begin{equation}
\beta(m, n) = \lambda_2 \int_{-\pi}^{\pi} d\phi \, \, e^{4 \sigma} 
J_2 (z_l) J_1(z_m) J_1(z_n),  
\label{beta}
\end{equation}
where 
\begin{equation}
\lambda_2 = m_m m_n \lambda_1.
\label{lambda2}
\end{equation}
In getting our results, we have ignored terms that come from the action of $\partial_\phi$ on the background
metric $G_{\mu \nu} = e^{-2 \sigma} \eta_{\mu \nu}$ or the prefactor $e^{-2 \sigma}$ of the 5-d graviton wavefunction in
Eq. (\ref{KKh}).  These terms, after integration by parts, are proportional to $(\sigma^\prime)^2$, where
$\sigma^\prime \equiv \partial_\phi \sigma$.  The terms proportional to $(\sigma^\prime)^2$ come with no
derivatives on the $\chi^{(n)}$-fields and thus would yield anomalous contributions to the 4-d Minkowski
couplings of the zero-mode graviton on the SM wall.  Hence, to maintain the RS construct, they must cancel
out, presumably after including the 5-d Cosmological Constant term contributions to the triple graviton
vertex.

\begin{figure}[htbp]
\centerline{
\psfig{figure=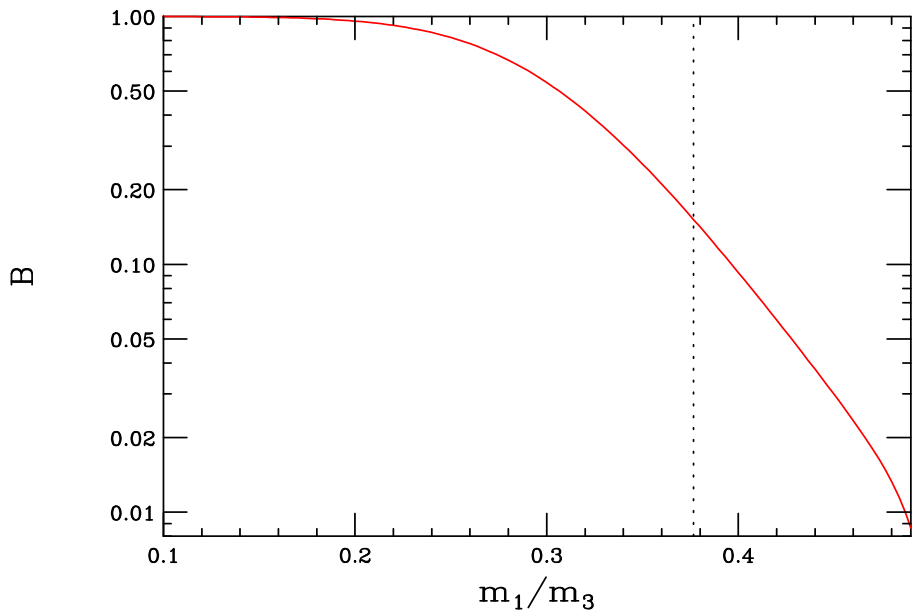,height=8cm,width=12cm,angle=90}}
\vspace*{0.5cm}
\caption[*]{Branching fraction for the decay $G^{(3)} \to G^{(1)}G^{(1)}$ as a function 
of the mass ratio $m_1/m_3$ assuming $m_3=2$ TeV. The fixed value of this 
ratio in the RS model is shown as the dotted line and yields $B \simeq 15\%$.}
\label{grav}
\end{figure}

Here, we note that the overall coefficient of the vertex $\{G^{(l)}, G^{(m)}, G^{(n)}\}$ has been fixed by
comparing the KK zero mode contribution to the vertex and matching with the 4-d results of Ref. \cite{BF},
after integrating over the extra dimension.  We may now numerically compute the decay rate for $\dcy$.  The
branching ratio for the decay as a function of $m_1/m_3$ is presented in Fig. (\ref{grav}).  Of course, at the
classical level in the RS model, $m_1/m_3 \simeq 3.83/10.17$, for which the branching fraction is about
$15\%$.  However, if the mass ratio is slightly different, perhaps due to quantum effects, the change in the
branching ratio is not negligible, as seen in  Fig. (\ref{grav}).  Higher KK modes may decay into more than two
$G^{(n)}$ resonances, \eg, $G^{(n+1)}\to nG^{(1)}$.  However, such decays come from Feynman diagrams
that are higher order in $\Lambda_\pi^{- 1}$.  In addition, perturbative calculations are unreliable at scales
much higher than $\lpi$.

In an AdS/CFT picture \cite{APR}, the KK states $G^{(n)}$ correspond to the resonances of a
strongly coupled 4-d CFT.  Thus, decays such as $G^{(3)} \to G^{(1)} G^{(1)}$ can be thought of
as the decays of heavier CFT resonances to lighter ones.  In the context of a CFT, there are $\sim
c^{-1}$ ``narrow''  resonances, where $c \equiv k/{\bar M}_P$, corresponding to the lightest
$G^{(n)}$ states, for $n \lsim c^{-1}$.  Again, the CFT states with masses much larger than
$\Lambda_\pi$ are too broad to be considered single resonances, and the computations based
on the particle picture of the $G^{(n)}$ are no longer adequate.  

\nn
\vspace*{0.4mm}

Here we note that the decay $\dcy$ can also get contributions from higher derivative operators
proportional to $R_5^2$, such as the 5-d Gauss-Bonnet term.  These contributions will be
parametrically non-leading, as the associated vertices are suppressed by $c^{-2/3}\Lambda_\pi^{- 3}$.  
One finds their relative contributions to the decay rate for $\dcy$ to be
$\sim (c^{1/3}x_3)^2$ where $x_3\simeq 10.1735$. Thus, even if $c \ll 1$ these terms may have a large
effect, assuming that their coefficients in the 5-d action are not too small.  This allows for the possibility
of measuring the 5-d $R_5^2$ coefficients of the theory at the $1\%$ level from measurements of
$G^{(n)}$ decays in future experiments, once the RS resonances are discovered.  In the case where the
$G^{(n)}$ have decays to three or more final state gravitons the presence of the $R_5^2$-like terms
will lead to a distortion of the final state spectrum thus providing a further handle on their relative
contribution beyond the total rate.

Probing the RS KK graviton sector and the bulk fermion mass matrix with precision, as discussed in this
and the next sections, requires large numbers of events.  Here, we note that a future LC will have ample
data to achieve such precise measurements.  To see this, consider the cross section for $e^+e^-\to
\mu^+\mu^-$ for the case where the SM is on the wall.  The branching
fraction for $\mu$ pairs is $\simeq 0.02$ and the light KK resonance 
 cross sections are $10^3-10^4 fb$ \cite{DHRprl}. 
 As the integrated luminosities of typical LC's are in the range of
$100-500~fb^{-1}/yr$, it is clear that one should easily expect $\sim 10^7$ graviton resonances in the
TeV range to be produced.  Similar arguments apply when the SM lies in the bulk and $\nu > - 0.5$ 
\cite{DHRprd}. 
When $\nu<-0.5$ one can instead use $\gamma \gamma$ collisions to produce graviton resonances at
rates of order $10^7/yr$.

\section{Flavor Changing Decays of Bulk Fields}

As described above, the large number of KK gravitons, $\sim 10^7$, produced at 
future lepton colliders will allow for a detailed study of their rare decay 
modes.  We now consider the case where the SM fields reside in the 5-d bulk.  
Since the parameter $\nu$ controls the strength of the 
couplings of both the 
graviton and gauge KK towers to the zero mode fermions, allowing fermions with 
the same SM quantum numbers to have different values of $\nu$ leads to 
Flavor-Changing (FC) couplings as noted by several 
authors {\cite {delA}.  The observation of rare decays induced 
by FC couplings could then be used to map out the 5-d fermion mass matrix.  
To be specific we consider a toy model where all of the fermions except the 
top quark have the same value of $\nu=-0.4$ and we let $\nu_t$ vary. (We choose 
the value $\nu=-0.4$ since it is middle of region III \cite{DHRprd} where all of 
the lightest graviton and gauge KK excitations are observable at colliders and 
there are no hierarchically large values of $\Lambda_\pi$.)  In this case, 
the mixing in the $Q=2/3$ sector will generate FC $u,c,t$ couplings which 
are somewhat less constrained than those in the $Q=-1/3$ sector \cite{Dphys}.  
Whether such FC couplings are sufficiently large as to be dangerous 
for low-energy $D$ meson physics is beyond the scope of the present paper.  In either 
case, the theory presented here is just a toy model for demonstration purposes 
and is not to be taken too seriously.

Let $c_i^{G,A}$ denote the couplings of a zero mode fermion, having $\nu=\nu_i$, 
to the lowest graviton or gauge excitation in the weak eigenstate basis, \ie, 
$c_i^{G,A}=C_{001}^{f\bar f G,A}(\nu_i)$ using the notation in Ref. \cite{DHRprd}.  (Here, 
$i=1, 2, 3$ will label the fields $u$, $c$ and $t$.)  Symbolically, we can write this interaction as 
$\bar f^{(0)}_i c_i^{G,A} f^{(0)}_i G^{(1)},A^{(1)}$. Converting to 
the mass eigenstate basis by 
a unitary transformation $U$ this becomes 
$\bar f^{(0)}_kU^\dagger_{ki}c_i^{G,A}U_{ij}f^{(0)}_iG^{(1)},A^{(1)}$.  Now consider the case $k=3$ and 
$j=1,2$ and use the fact that $c_1^{G,A}=c_2^{G,A}$ together with the 
unitarity condition  
$\sum_i U^\dagger_{ki}U_{ij}= \delta_{k j}$; the strength of the FC couplings are found to 
be $(c_3^{G,A}-c_1^{G,A})U^\dagger_{33}U_{31,2}$.  The difference in the $c_i$ 
can be obtained directly by performing 
some basic integrations over Bessel functions, once the value of $\nu_t$ is specified.  
To obtain sample numerical results 
we will assume that $U_{ij}\sim (V_{CKM})_{ij}$ at least qualitatively 
so that, \eg , $U_{33}=1$ and 
$U_{32}\simeq A\lambda^2\simeq 0.04$.  Note that in this toy model the $t\bar c$ 
coupling is thus expected to be an order of magnitude larger than the corresponding $t\bar u$ one. 

\vspace*{-0.5cm}
\nn
\begin{figure}[htbp]
\centerline{
\psfig{figure=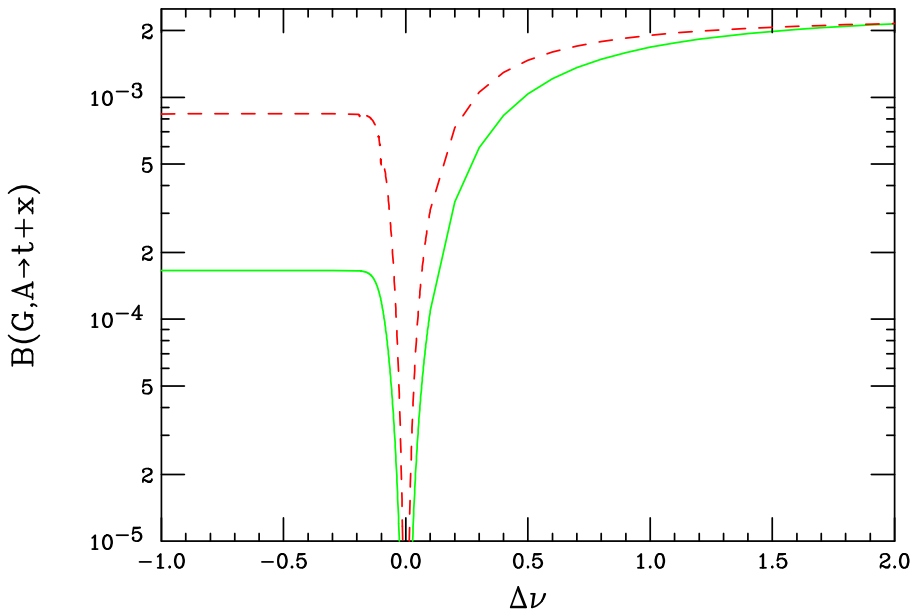,height=8.0cm,width=12cm,angle=90}}
\vspace*{0.1cm}
\caption[*]{The flavor changing branching fractions for a 1 TeV first 
graviton (solid green) or $\gamma/Z$ (dashed red) excitation as functions of 
$\Delta \nu$.}
\label{fcnc}
\end{figure}
\vspace*{0.4mm}

It is now straightforward to obtain the desired branching fractions $B$ for 
$G,A\to t\bar x+x\bar t$ which are shown in Fig. (\ref{fcnc}) as functions of 
$\Delta \nu=\nu_t-\nu$ with $\nu=-0.4$ assuming first KK masses of 1 TeV.  (For 
the gauge boson KK case we assume that we are probing the almost degenerate 
first $\gamma/Z$ excitation.  Comparable results would be obtained for the 
first KK gluon excitation.)  We see immediately that if 
$\Delta \nu$ is sizeable, \eg \, $\Delta \nu \grtsim 0.1$, branching fractions as large as 
$10^{-(3-4)}$ are obtainable.  Decays at such high rates will be trivial to 
observe with a sample of $\sim 10^7$ gravitons or gauge KK excitations.  
For very small $\Delta \nu$ we find the approximate relations 
$B\simeq 1.4 \times 10^{-6} [\Delta \nu/0.01]^2$ for gravitons and 
$B\simeq 5.3 \times 10^{-6} [\Delta \nu/0.01]^2$ for neutral gauge bosons. Using 
these results it is clear that values of $|\Delta \nu| \sim 0.01$
can be probed through these decays, at LC Graviton Factories.

\section{Conclusion}

In this paper we have explored some of the physics accessible at LC Graviton 
Factories which may produce as many as $10^7$ resonant gravitons per 
year at design luminosities, depending upon their mass.  We 
examined two particular processes: ($i$) the decay of a heavy graviton tower 
member into two lighter ones and ($ii$) the flavor changing decay of a light 
graviton (or bulk vector field) when the the SM is off the wall. 

As described in the text, the tree-level coupling of gravitons to matter occurs at the same order as does the
trilinear graviton coupling, in analogy with QCD.  In order to explore the underlying theory of gravity in the RS
model, it is necessary to probe these gravitational self-couplings.  This becomes possible in the RS model at
Graviton Factories where we can observe the decay of higher KK states into lighter ones.  We note that by
the AdS/CFT correspondence, one could interpret these KK gravitons as the resonances of a strongly
coupled 4-d CFT \cite{APR}.  In particular, we examined the process $G^{(3)} \to G^{(1)} G^{(1)}$ which due to the RS
particle spectrum is the first accessible one at colliders.  The branching fraction for this decay was found to
be about $15\%$.  This branching fraction yields a precise measurement of the gravitational self-interactions
which allows us to probe, \eg, higher derivative contribution in the 5-d action, such as the Gauss-Bonnet
term, at the level of $1\%$ or better.  The decay rate shows sensitivity to $m_1/m_3$; this ratio could in
principle receive quantum corrections.

In the case where the SM fields lie in the bulk, giving fermions of each  
generation with the same electroweak charges different values of the 
parameter $\nu$ leads to FCNC-type decays.  If the branching fractions for 
these decays are precisely measured we can map out the values of $\nu$ for all 
the fermions.  In this paper, we demonstrated, within the context of a toy model, 
that such FCNC's can be sizeable for both graviton and gauge KK resonances 
and that the precision measurements at a Graviton Factory 
can probe differences in $\nu$ values as low as 0.01.  Together 
with ordinary flavor conserving decays, these can be used to map out all of the 
fermion locations in the extra dimension.

\noindent{\Large\bf Acknowledgements}

We would like to thank D.J. Chung, J.L. Hewett, and M.E. Peskin for discussions.

\newpage

%
\def\MPL #1 #2 #3 {Mod. Phys. Lett. {\bf#1},\ #2 (#3)}
\def\NPB #1 #2 #3 {Nucl. Phys. {\bf#1},\ #2 (#3)}
\def\PLB #1 #2 #3 {Phys. Lett. {\bf#1},\ #2 (#3)}
\def\PR #1 #2 #3 {Phys. Rep. {\bf#1},\ #2 (#3)}
\def\PRD #1 #2 #3 {Phys. Rev. {\bf#1},\ #2 (#3)}
\def\PRL #1 #2 #3 {Phys. Rev. Lett. {\bf#1},\ #2 (#3)}
\def\RMP #1 #2 #3 {Rev. Mod. Phys. {\bf#1},\ #2 (#3)}
\def\NIM #1 #2 #3 {Nuc. Inst. Meth. {\bf#1},\ #2 (#3)}
\def\ZPC #1 #2 #3 {Z. Phys. {\bf#1},\ #2 (#3)}
\def\EJPC #1 #2 #3 {E. Phys. J. {\bf#1},\ #2 (#3)}
\def\IJMP #1 #2 #3 {Int. J. Mod. Phys. {\bf#1},\ #2 (#3)}

\end{document}